\documentclass[aps,prd,twocolumn,nofootinbib,groupedaddress,amsfonts,floatfix]{revtex4}

\usepackage{graphicx,amsmath,amssymb,amstext}
\usepackage{amssymb,amsbsy,amsfonts,amsthm,hyperref}
\usepackage{lscape,placeins}
\usepackage{cancel}
\usepackage{appendix}
\usepackage[utf8]{inputenc}
\usepackage{multirow}

\newcommand{\lsim}{\;\mbox{\raisebox{-0.5ex}{$\stackrel{<}{\scriptstyle{\sim}}$}}\;}

\usepackage{color}
\usepackage{ifthen}
\newboolean{editorial}
\setboolean{editorial}{true}
\newcommand{\editorial}[2]{\ifthenelse{\boolean{editorial}}{\textcolor{red}{[\textsf{\textbf{{#1}}}: }\textcolor{blue}{\textsf{{#2}}}\textcolor{red}{]}}{}}

\begin{document}

\title{Observable Deviations from Homogeneity in an Inhomogeneous Universe}

\author{John T. Giblin, Jr${}^{1,2}$}
%\email[]{giblinj@kenyon.edu}
\author{James B. Mertens${}^{2}$}
%\email[]{jbm120@case.edu}
\author{Glenn D. Starkman${}^{2}$}
%\email[]{glenn.starkman@case.edu}

\affiliation{${}^1$Department of Physics, Kenyon College, 201 N College Rd, Gambier, OH 43022}
\affiliation{${}^2$CERCA/ISO, Department of Physics, Case Western Reserve University, 10900 Euclid Avenue, Cleveland, OH 44106}

\begin{abstract}
How does inhomogeneity affect our interpretation of cosmological observations?
It has long been wondered to what extent the observable properties of an
inhomogeneous universe differ from those of a corresponding
Friedman-Lemaitre-Robertson-Walker (FLRW) model, and how the inhomogeneities
affect that correspondence. Here, we use numerical relativity to study the
behavior of light beams traversing an inhomogeneous universe and construct
the resulting Hubble diagrams. The universe that emerges exhibits an average
FLRW behavior, but inhomogeneous structures contribute to deviations in
observables across the observer's sky. We also investigate the relationship
between angular diameter distance and the angular extent of a source, finding
deviations that grow with source redshift. These departures from FLRW are
important path-dependent effects with implications for using real observables
in an inhomogeneous universe such as our own.
\end{abstract}

\maketitle

\section{Introduction}

The field of cosmology has always relied on the cosmological
principle: the Universe is homogeneous and isotropic. 
This principle has been used to construct the dynamical models with
which we understand and interpret the evolution of the Universe; and in the
context of this principle we define and measure cosmological parameters.
However, as cosmological measurements reach percent-level precision, 
the effects of inhomogeneities in the Universe become increasingly
important when comparing observations to models. 
The field of cosmology has typically studied these effects using
a perturbative approach, with inhomogeneities described and modeled by
a linearized-gravity approximation \cite{Bernardeau:2001qr,Adamek:2015eda}.
Additionally, semi-analytic models have been used to study effects of inhomogeneities on
observables in specific spacetimes, such as Swiss Cheese or Lindquist-Wheeler
metrics \cite{Fleury:2013sna,Fleury:2014gha,Bolejko:2012ue,Clifton:2011mt}.

Just a short time ago, the dynamics of cosmological dust spacetimes were examined
in full 3+1-dimensional general relativity (GR) for the first time by solving the Einstein
field equations numerically \cite{Giblin:2015vwq,Mertens:2015ttp,Bentivegna:2015flc}.
These studies and others such as \cite{Yoo:2013yea,Bentivegna:2013jta,Yoo:2014boa} have, 
so far, focused on mathematical and numerical properties of the spacetimes,
rather than the impact of inhomogeneities on cosmological observables. 
An open question therefore remained: can we compute predictions for observables
and demonstrate the necessity and power of these numerical simulations?

In this work, we present the first calculation of observable quantities in cosmology
using the full, unconstrained framework of numerical relativity for a pure ``dust''
spacetime containing inhomogeneities on large scales -- a framework consistent
with the work presented in \cite{Giblin:2015vwq,Mertens:2015ttp,Bentivegna:2015flc}.
We follow light beams along null geodesics through a toy universe and integrate the
optical scalar equations along these geodesics, garnering information about angular
diameter distances and photon redshifts as the universe evolves. Using these
observables, we construct the resulting Hubble diagrams, plotting the distance
modulus versus redshift, and compare the results to the
Friedman-Lemaitre-Robertson-Walker (FLRW) model.

\begin{figure}[hb!]
  \centering
    \includegraphics[width=0.45\textwidth]{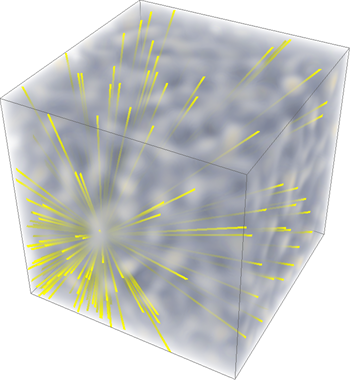}
  \caption{\label{raytracing} A depiction of raytracing through a spacetime with
  density fluctuations. Shades of lighter gray represent denser regions of
  space, and darker colors represent less dense regions. In a time-reversed
  sense, light rays (yellow) originate from an observer located in the volume
  and propagate outward along null geodesics.}
\end{figure}

We thus extend the previous numerical studies in GR by examining the properties of
photons traversing these spacetimes. A schematic of this scenario is depicted in
Fig.~\ref{raytracing}. In general, we find good agreement between the averaged
observables and corresponding FLRW model; however, there are departures along
individual paths, and when observers or sources are located in under-densitites
or over-densities. 

The demonstration of this method by way of drawing Hubble diagrams 
is the main result of this work. We explore the ability
of our code to resolve effects that manifest themselves in Hubble diagrams
due to observers being located in over-dense or under-dense regions
\cite{Rees:1968zza,GarciaBellido:2008nz,Biswas:2010xm}.
We additionally demonstrate the ability of this method to examine the
relationship between different distance measures, such as luminosity distance
and observed angular extent, probing discrepancies that might arise due to 
inhomogeneities. While our model attempts to mimic our Universe on large
scales, we do not yet resolve small-scale structure, and that the amplitude
of fluctuations is not observationally motivated. Nevertheless, we hope to
be able to model these more realistically in the future.

We begin in Section~\ref{bssn_eoms} by briefly reviewing the formulation of
numerical relativity used in this work, and describing, in
Sections~\ref{DA_eom} and ~\ref{geodesic_eom}, the formalism we use to
integrate photon geodesics. We then provide details of the numerical technique
used to integrate these equations in Section~\ref{num_ev}.
Finally, in Section~\ref{results_sec} we present results
from a series of cosmologically-motivated simulations, and remark
on Etherington's reciprocity relation and the ability of our
technique to probe outstanding questions in cosmology.

\section{Evolution of observables in a general spacetime}

Numerical relativity has seen a number of breakthroughs in the past two decades, 
notably the development of formulations capable of stable long-term evolution. 
We take advantage of these advancements, and once again \cite{Mertens:2015ttp}
apply them in a cosmological setting. We also write the geodesic equations and
optical-scalar equations in a form suitable for integration in our numerical
spacetime.

\subsection{The BSSN formulation} \label{bssn_eoms}

The BSSN formulation \cite{Baumgarte:1998te,Shibata:1995we} is a variant of a
$3+1$ decomposition with important modifications for improving the numerical
stability of the system. For a review or for further details, see eg.
\cite{Cardoso:2014uka,BaumgarteShapiroBook}. In
this formulation, the metric is decomposed as
\begin{eqnarray}
g_{\mu\nu} & = & \left(\begin{array}{cc}
-\alpha^{2}+\beta_{l}\beta^{l} & \beta_{i}\\
\beta_{j} & e^{4\phi}\bar{\gamma}_{ij}
\end{array}\right)\,,
\end{eqnarray}
where $\gamma_{ij}=e^{4\phi}\bar{\gamma}_{ij}$ is the spatial metric with
$\det(\bar{\gamma}_{ij})=1$, and $\alpha$ and $\beta^{i}$ are the lapse and
shift respectively. The metric obeys the BSSN evolution equations,
\begin{eqnarray}
\label{BSSN_equs}
\partial_{t}\phi & = & -\frac{1}{6}\alpha K+\beta^{i}\partial_{i}\phi+\frac{1}{6}\partial_{i}\beta^{i}\\
\partial_{t}\bar{\gamma}_{ij} & = & -2\alpha\tilde{A}_{ij}+\beta^{k}\partial_{k}\bar{\gamma}_{ij}+\bar{\gamma}_{ik}\partial_{j}\beta^{k} \nonumber\\
& & +\bar{\gamma}_{kj}\partial_{i}\beta^{k}-\frac{2}{3}\bar{\gamma}_{ij}\partial_{k}\beta^{k}\\
\partial_{t}K & = & -\gamma^{ij}D_{j}D_{i}\alpha+\alpha(\tilde{A}_{ij}\tilde{A}^{ij}+\frac{1}{3}K^{2})\nonumber\\ 
 & & +4\pi\alpha(\rho+S)+\beta^{i}\partial_{i}K
\end{eqnarray}
\begin{eqnarray} %% -- looks better with a column break here
\partial_{t}\tilde{A}_{ij} & = & e^{-4\phi}(-(D_{i}D_{j}\alpha)+\alpha(R_{ij}-8\pi S_{ij}))^{TF}\nonumber\\ 
 & & +\alpha(K\tilde{A}_{ij}-2\tilde{A}_{il}\tilde{A}_{j}^{l}) +\beta^{k}\partial_{k}\tilde{A}_{ij} \nonumber \\
 &  & +\tilde{A}_{ik}\partial_{j}\beta^{k}+\tilde{A}_{kj}\partial_{i}\beta^{k}-\frac{2}{3}\tilde{A}_{ij}\partial_{k}\beta^{k}\,,
\end{eqnarray}
with source terms $\rho$, $S$, $S_{j}$, and $S_{ij}$ projections of the
stress-energy tensor onto the spatial slice.
A key to the numerical stability of the system is the introduction of a
redundant set of auxiliary variables, contractions of the Christoffel symbol of the
conformal metric $\bar{\gamma}_{ij}$,
$\bar{\Gamma}^{i} \equiv \bar{\gamma}^{jk}\bar{\Gamma}_{jk}^{i}$, which are used
to compute the Ricci tensor. These are also dynamically evolved,
\begin{equation}
\partial_{t}\bar{\Gamma}^{i}=2\left(\bar{\Gamma}_{jk}^{i}\tilde{A}^{jk}-\frac{2}{3}\bar{\gamma}^{ij}\partial_{j}K-8\pi\bar{\gamma}^{ij}S_{j}+6\tilde{A}^{ij}\partial_{j}\phi\right)\,.
\end{equation}
In this scheme, the extrinsic curvature, $K_{ij}$, has been decomposed into a
conformally related trace-free part, $\tilde{A}_{ij}$, and trace, $K$, as
$K_{ij}=e^{4\phi}\tilde{A}_{ij}+\frac{1}{3}\gamma_{ij}K$.
Although our code allows for an arbitrary gauge choice, synchronous gauge
($\alpha=1$ and $\beta^{i}=0$) is applied in this work.

A pressureless, $w=0$, perfect fluid is used to model the matter component of
this universe on large scales, consistent with a cold-dark matter model.
The general equations of motion for such a fluid require integration techniques
tailored to the problem, such as finite-volume methods \cite{10.2307/41149030},
phase-space methods \cite{1989ApJ...344..146R}, or N-body techniques
\cite{Olabarrieta:2000bn}. However, in synchronous gauge, the equations of
motion simplify to
\begin{equation}
\partial_t (\gamma^{1/2}\rho) = 0
\end{equation}
for a fluid with no initial coordinate velocity and density $\rho$. Although
multistreaming -- or the crossing of fluid elements -- cannot be resolved using
this formulation due to the formation of coordinate singularities, this does not
present an issue on the large scales examined in this work. We can therefore
choose an initial conformal density $\gamma^{1/2}\rho$ that will not evolve over
the course of the simulation. For this choice of coordinates, the remaining
matter source terms in the BSSN Equations are $S = S_{j} = S_{ij} = 0$.

\subsection{Propagation along null geodesics} \label{geodesic_eom}

While the BSSN formalism allows us to evolve the metric, we wish also to compute
measurable quantities. We therefore derive the equations of motion for particles
traveling along geodesics. Although we are particularly interested in null
geodesics, the equations presented in this section can also be applied to massive
particles.

Adopting the notation of (and loosely following) \cite{Vincent:2012kn}, we
begin by noting that the equations describing propagation along a
geodesic are often parametrized by an affine variable, $\lambda$. We will need to
re-parametrize these equations in terms of the coordinate time $t$ and variables
from a 3+1 decomposition. The 4-momentum, $p^{\mu}$, for a particle following a
geodesic is
\begin{equation}
\frac{{\rm d}X^{\mu}}{{\rm d}\lambda}=p^{\mu}\,,
\end{equation}
for an affine parameter $\lambda$ that parametrizes the particle's path.
In order to work in terms of $3+1$ quantities, the 4-momentum is decomposed and
written in terms of quantities according to observers on a spatial slice. This
is accomplished by writing the 4-momentum in terms of a piece parallel to the
unit normal to the slices, $p_{\parallel}^{\mu} \parallel n^\mu$, where the
normal is $n^{\mu} = (\alpha^{-1},-\alpha^{-1}\beta^{i})$, as well as a transverse
piece proportional to the velocity according to a normal observer,
$p_{\perp}^{\mu} \propto V^\mu$.

We then define the energy of the particle according to a normal observer as
$E \equiv -n_{\mu}p^{\mu}$. The zero-component of the 4-momentum can be related
to $E$ by noticing that
\begin{equation}
n_{\mu}{\rm d}X^{\mu}=n_{\mu}p^{\mu}d\lambda,
\end{equation}
which is just $E/\alpha=p^{0}$. We can now decompose the 4-momentum as
\begin{eqnarray}
p_{\parallel}^{\mu} & = & p^{\mu}-p_{\perp}^{\mu}=En^{\mu} \nonumber \\
p_{\perp}^{\mu} & = & p^{\mu}-En^{\mu}\equiv EV^{\mu}\,,
\end{eqnarray}
where $p^{\mu}=E\left(n^{\mu}+V^{\mu}\right)$. Because the velocity vector
$V^\mu$ has no component in the direction normal to the spatial slices
($n_{\mu}V^{\mu}=0$), 
we can choose to write it as $V^{\mu}=(0,V^{i})$, with no time
component.

We now wish to obtain evolution equations for $E$ and
$V^i$ in terms of a coordinate time $t$ rather than affine parameter.
We can do so using the geodesic equation,
\begin{equation}
\frac{{\rm d}^{2}X^{\mu}}{{\rm d}\lambda^{2}}=-\Gamma_{\alpha\beta}^{\mu}\frac{{\rm d}X^{\alpha}}{{\rm d}\lambda}\frac{{\rm d}X^{\beta}}{{\rm d}\lambda}
\end{equation}
for the position of a particle, $X^{\mu}$, along a path parametrized by
$\lambda$. This can be written in a form well-suited for integrating the $3+1$
variables, resulting in a closed set of equations of motion:

\begin{eqnarray}
\frac{1}{E}\frac{{\rm d}}{{\rm d}t}E & = & \alpha K_{ij}V^{j}V^{k}-V^{j}\partial_{j}\alpha\\
\frac{{\rm d}X^{i}}{{\rm d}t} & = & \alpha V^{i}-\beta^{i}\label{eq:Xev}\\
\frac{{\rm d}V^{i}}{{\rm d}t} & = & \alpha V^{j}\left(V^{i}\partial_{j}\ln\alpha-K_{jk}V^{k}V^{i}+2K_{j}^{i}\right.\nonumber\\
 & & \left.-{}^{(3)}\Gamma_{jk}^{i}V^{k}\right)-\gamma^{ij}\partial_{j}\alpha-V^{j}\partial_{j}\beta^{i}\,.\label{eq:Vev}
\end{eqnarray}

\subsection{Angular diameter distances and the optical scalar equations}
\label{DA_eom}

The angular diameter distance, $D_{A}$, is a standard distance measure used in
cosmology, with the desirable property that it can be directly measured or
computed for distant sources. It is also related to the luminosity distance:
$D_{A}=(1+z)^{2}D_{L}$. This statement, known as the Etherington Reciprocity
Theorem or distance duality relationship \cite{Etherington2007,Ellis2007},
is a geometric result, and holds so long as photon number is conserved and all
photons in the beam ``feel'' the same metric -- meaning the beam width is small
compared to distances over which the metric varies. This small beam limit should
be a good approximation for point-like sources such as supernovae, but not
necessarily for beams with a large angular extent, such late-universe baryon
acoustic oscillations, or degree-scale CMB fluctuations. A more thorough
discussion can be found in other works, eg. \cite{Seitz:1994xf}.

In order to compute angular diameter distances, we must track the
cross-sectional area of a light beam as it propagates through space. We
accomplish this using the Sachs optical-scalar equations, rewritten in a form
well-suited to both numerical integration and extraction of cosmological
parameters. These equations track the width of a beam $\ell$, its
time-derivative $\varphi$, the shear rate of the beam $\sigma$, and basis
vectors, known as screen vectors, spanning the plane of the beam area, $s_1^\mu$ and
$s_2^\mu$. The evolution equations (briefly elaborated upon in
Appendix~\ref{angular_diameter_appendix} and Appendix~\ref{weyl_calcs}) are
\begin{eqnarray}
p^{0}\frac{{\rm d}}{{\rm d}t}\varphi & = & \ell\left(\mathcal{R}-\sigma_{R}^{2}-\sigma_{I}^{2}\right)\nonumber\\
p^{0}\frac{{\rm d}}{{\rm d}t}\ell & = & \varphi\nonumber\\
p^{0}\frac{{\rm d}}{{\rm d}t}\bar{\sigma}_{R} & = & \ell^{2}\Re[\mathcal{W}]\nonumber\\
p^{0}\frac{{\rm d}}{{\rm d}t}\bar{\sigma}_{I} & = & \ell^{2}\Im[\mathcal{W}]\,,
\end{eqnarray}
with $\bar{\sigma}=\ell^{2}\sigma$. The Ricci optical scalar $\mathcal{R}$ is
given by
\begin{eqnarray}
\mathcal{R} & = & -4\pi T_{\mu\nu}p^{\mu}p^{\nu}\,,
\end{eqnarray}
and the real and imaginary pieces of the Weyl scalar $\mathcal{W}$ are
\begin{eqnarray}
\Re[\mathcal{W}] & = & -\frac{1}{2}C_{\mu\nu\rho\sigma}(s_{1}^{\mu}s_{1}^{\sigma}-s_{2}^{\mu}s_{2}^{\sigma})p^{\nu}p^{\rho}\\
\Im[\mathcal{W}] & = & C_{\mu\nu\rho\sigma}s_{1}^{\mu}p^{\nu}p^{\rho}s_{2}^{\sigma}\,.
\end{eqnarray}

The screen vectors evolve according to
\begin{eqnarray}
\frac{\rm{d}}{\rm{d}t}s_A^\mu & = & s_{A}^{j}V^{i}\left(\gamma_{jk}\frac{\rm{d}}{\rm{d}t}V^{k}-3V^{k}K_{jk}+V_{l}V^{k}\Gamma_{kj}^{l}\right) \nonumber \\
 & & -\Gamma_{kl}^{i}V^{k}s_{A}^{l}+\gamma^{ij}K_{jk}s_{A}^{k} \,.
\end{eqnarray}
These equations are subject to boundary conditions at the point of
observation, namely that the beam area has converged, $\ell(t_{obs})=0$,
the observer views the beam as subtending some solid angle
$\varphi(t_{obs})=\sqrt{\Omega_{obs}}$, and the shear rate is $\sigma=0$
\cite{Fleury:2013sna}. In terms of these variables, the angular diameter
distance to a source that emitted light at some time $t_{em}$,
seen by an observer as subtending a solid angle $\Omega_{obs}$ at time $t_obs$,
is $D_A = \ell(t_{em}) / \varphi(t_{obs})$.

We conclude this section by writing the relationship between various quantities
defined so far, and corresponding FLRW quantities for a completely homogeneous
spacetime, in Table~\ref{frw_ref_table}.

\begin{table}[h]
\centering
\begin{tabular}{c|c|c}
Quantity & FLRW Notation & Our Notation \\
\hline
Scale factor & $a$ & $e^{2\phi}$ \\
Hubble parameter & $H$ & $-K/3$ \\
Photon redshift & $z$ & $E-1$ \\
Angular diameter distance & $D_A$ & $\ell_{em} / \varphi_{obs}$  \\
\end{tabular}
\caption{Translation between variables in a completely homogeneous dust universe.}
\label{frw_ref_table}
\end{table}

\section{Numerical Method Details} \label{num_ev}

As we are interested in examining the dynamics of photon geodesics
converging at an observer, we will need to ensure our boundary conditions
at the observer are satisfied, or that beams have converged at a particular
spacetime point. The most straightforward and consistent way to implement this
is to integrate a system forward from some initial state to a future
observation time, and then trace photon paths while integrating the system of
equations backwards in time. Our code implements precisely this, allowing us to
evolve the system from some cosmological initial conditions to a future state,
change the sign of the timestep, and continue integrating both the metric and
photon geodesics. The cosmological initial conditions we use consist of a
Gaussian-random realization of a matter field with a cosmologically-motivated
matter power spectrum in a periodic spacetime,
\begin{equation}
P_k^{\rho} = \frac{4P_*}{3} \frac{ k/k_*}{1+\frac{1}{3}(k/k_*)^{4}}\,.
\end{equation}
We cut $P_k^{\rho}$ off at some finite $k_{\rm cutoff}$ in order to reduce
fluctuations on
scales where grid effects become important, so that fluctuations are resolved
by sufficiently many points. Further details of this spacetime can be found in
\cite{Mertens:2015ttp}. In order to specify initial conditions for light rays,
we pick random outgoing directions from a specified point, and integrate along
null geodesics in these directions backwards in time.

Integrating backwards not only allows us to ensure light rays converge at
an observer, but also allows us to verify that we evolve back to the initial
state imposed at the beginning of the simulation. We find that our code passes
this check to within a small, expected level of error.
While of importance in our work, such a technique would not necessarily be
suitable for simulations without periodic boundary conditions, as information
about the state of the metric may propagate outside the volume of interest when
using, for example, damping boundary conditions. This method may also fail for
hydrodynamical calculations in the presence of shocks.

Because we integrate through a discretized spacetime, we also need
to compute metric quantities between grid points, requiring an interpolation
scheme. In this code we utilize simple linear interpolation, accurate only to
$\mathcal{O}(\Delta x^{2})$. In principle this could be improved, as the metric
and matter fields are evolved using an $\mathcal{O}(\Delta x^{8})$ scheme,
however this level of accuracy is not found to be necessary here.
Accumulated error for quantities integrated along geodesics is therefore
$O(\Delta x)$, the dominant contribution to the error in our results.
For cosmological runs (such as presented in \cite{Giblin:2015vwq}), we
see convergence as expected given the interpolation and integration order
used to follow photon trajectories. Further details of the code we use to
evolve the spacetime can be found in
\cite{Mertens:2015ttp}.

We use results from simulations run at three different resolutions:
$N^3 = 128^3$, $N^3 = 144^3$ and $N^3 = 160^3$. We then use a standard
numerical technique, Richardson extrapolation
\cite{Press:2007:NRE:1403886,Alcubierre:1138167}, in order to test convergence,
to obtain results at higher order, and to obtain error estimates.

As a final note on error, the statistical uncertainty in computed means scales
as $1/\sqrt{N_\gamma}$, with $N_\gamma$ the number of traced rays. In
general, we will be tracing $N_\gamma = 10^3$ rays for each observer, so we
expect statistical error in the means to be a few per-cent of the variance.
Numerical error in computed quantities is found to be of order a part in
$10^6$, significantly smaller than this statistical error.

\section{Results: Computing Observables} \label{results_sec}

In this section we present results from following light beams through a series
of simulated cosmological spacetimes similar to the fiducial model presented in
\cite{Giblin:2015vwq}. We begin by examining the properties of observables in our
model, and describe the effects of the inhomogeneities compared to a corresponding
FLRW model, or a homogeneous universe containing only pressureless matter
(eg., $\Omega_m = 1$). The initial density fluctuations in these runs are around
$\sigma_\rho/\rho = 0.1$, which grow to $\sigma_\rho/\rho = 0.2$ on the spatial slice
containing our observers. On average, we see that our simulations reproduce the
FLRW universe extremely well; however there are significant deviations along
individual paths and when observers or sources are preferentially located in
under- or over-dense regions.

\begin{figure}[htb]
  \centering
    \includegraphics[width=0.45\textwidth]{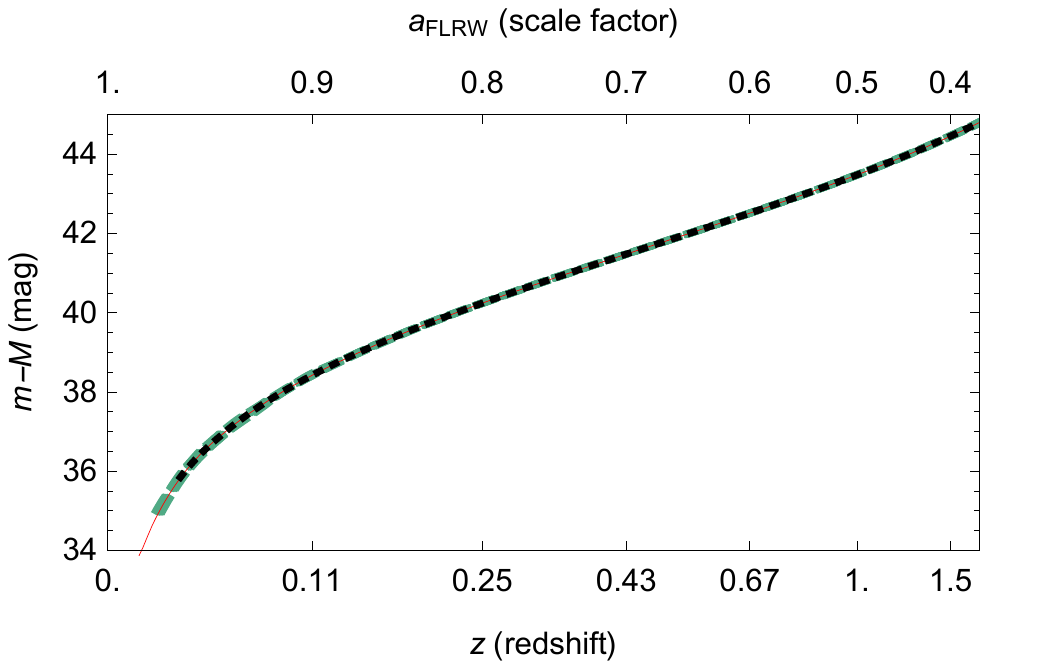}
  \caption{\label{avg_hubble}
  A Hubble diagram according to an `average' observer. The black dotted line
  indicates the prediction from a pure-FLRW model, many green dashed lines are
  derived from angular diameter distances computed along individual lines of
  sight, and the red line is the average along all individual lines of sight. The
  different lines are nearly indistinguishable.
  }
\end{figure}

We first remark on the case of an observer whose local density corresponds
closely to the average value on a spatial slice. In this case, we find light
beams traversing individual geodesics trace out a uniform distribution of
angular diameter distances, the mean of which closely tracks the FLRW expectation
at all redshifts we examine. In Fig.~\ref{avg_hubble} we present a Hubble
diagram as would be computed by such an average observer, after converting the
angular-diameter distances to apparent magnitudes,
$m-M = 5\log_{10}\left(D_A (1+z)^{-2}/10\,{\rm pc}\right)$.

\begin{figure}[t]
  \centering
    \includegraphics[width=0.45\textwidth]{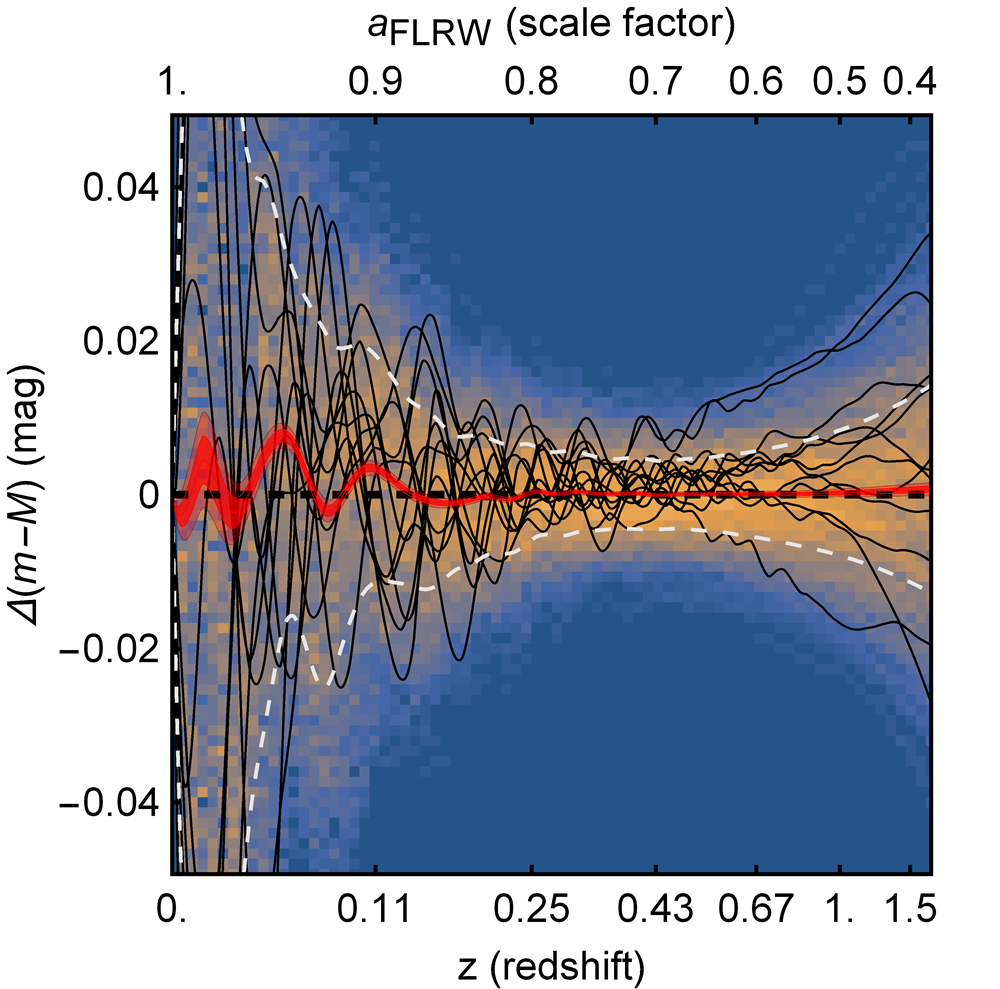}
  \caption{\label{avg_hubble_dev}
  A residual Hubble diagram according to an `average' observer.
  Thin black lines depict angular diameter distances along several particular
  lines of sight for this observer. The black dashed line indicates the prediction
  from a pure-FLRW model. The solid red line indicates the average angular diameter
  distance of these rays, for which we see agreement with FLRW. The statistical
  error -- standard error in the mean -- is shown at the one and two `sigma` level,
  indicated by semi-transparent red shading, so the width of the red curve is
  indicative of error. White dashed lines indicate the standard deviation of the
  distribution of magnitudes. Finally, the background shows a histogram of the
  magnitudes along all integrated lines of sight as a function of redshift.
  }
\end{figure}

In Fig.~\ref{avg_hubble_dev}, we present the corresponding residual Hubble
diagram.  In this diagram we subtract the distance modulus for a matter-dominated
FLRW model from the distance modulus we calculate for our model for each value of $z$.
Of note, there is some ambiguity in choosing which FLRW model to compare to.
For example, the FLRW model that agrees with the conformal average density
at a particular time will not necessarily agree at a later time, and will also
not necessarily agree with the FLRW model that agrees with the (conformal)
average expansion rate. For an increasingly inhomogeneous universe these
differences may become important, however for our simulations, the differences
are largely unimportant. We thus compare to a pure-FLRW model for which the
expansion rate agrees with the conformally averaged expansion rate of the
initial simulation hypersurface.

In the residual diagram, we find that despite fluctuations along
individual paths, there is close agreement between the average angular diameter
distance and corresponding FLRW model at most of the redshifts we probe.
Of significance, at very low redshift ($z \lsim 0.1$), there are sizable
deviations. These arise primarily due to the local structure of the universe:
although this observer is situated at a location of near-average density, they
are nevertheless near both over-dense and under-dense patches, which in turn
affect angular-diameter measurements. Averaged across many such observers, these
local effects can disappear; however for an individual observer we do not find
this to be the case. At larger redshifts, we find agreement between averaged
quantities and the corresponding FLRW model to within the uncertainty of our model.

We could compare this behavior to semi-analytic models, for example Swiss-Cheese
universes, using the Dyer-Roeder approximation. To an extent, our findings
contrast the ideas behind such models, which posit that light beams will undergo
less focusing as they intercept less matter (for a given average density) in
``lumpier" universes. However, our current simulations are in a regime that should be
well-described by linear theory, or a regime in which fluctuations are normally
distributed around the FLRW solution, and thus average away. These models also
suggest the shear terms are negligible, something we find approximate agreement with;
the size of the shear terms relative to the Ricci optical scalar is
$\mathcal{R}/|\sigma^2| \sim \mathcal{O}(0.01\%)$. Nevertheless, once collapsed
structure is better resolved by future work, investigating these comparisons
will be an interesting and important task.

\begin{figure}[t]
  \centering
    \includegraphics[width=0.45\textwidth]{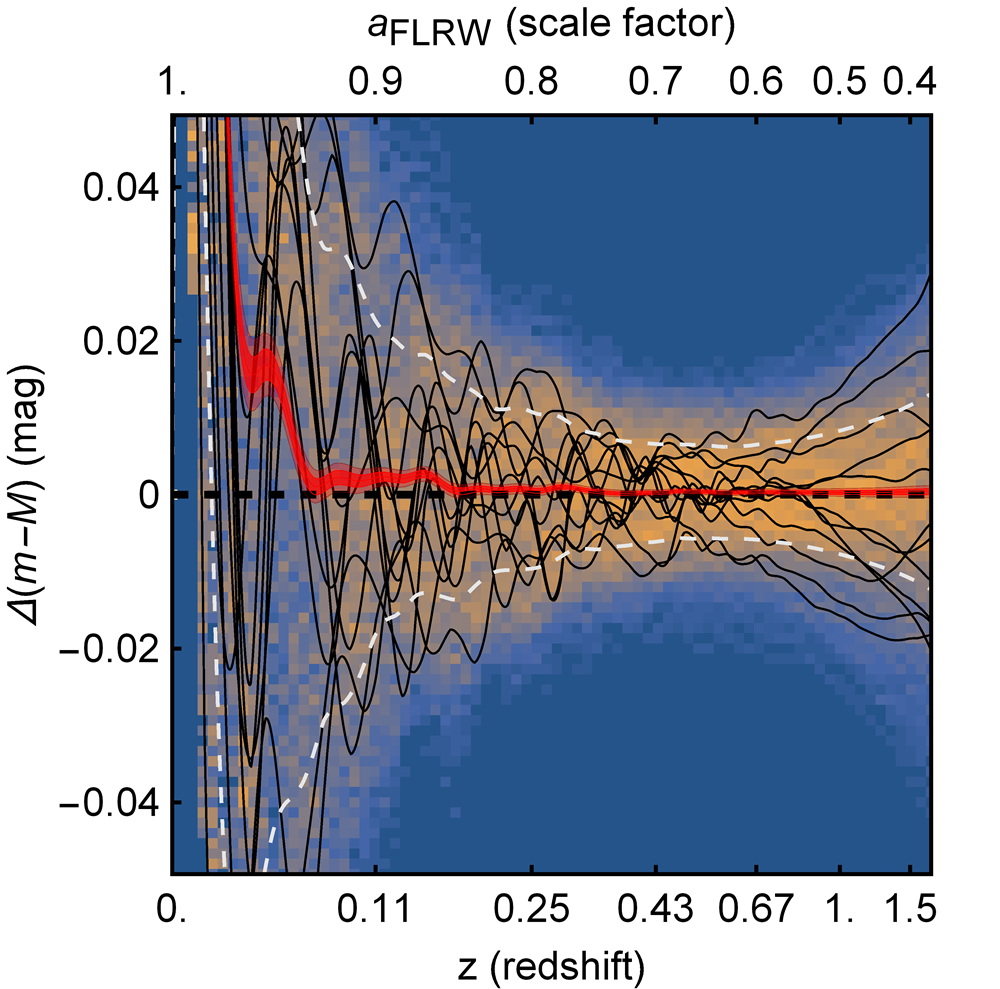}
  \caption{\label{Hubble_overdense} A Hubble diagram as in figure \ref{avg_hubble_dev}
  created by raytracing through the simulated universe, but for an observer located
  in an overdensity. Effects due to local structure are manifest at low redshift. }
\end{figure}

We next examine effects due to an observer being situated in a local over-dense or
under-dense region.
In synchronous gauge, especially at early times, fluctuations in the metric are directly
sourced by fluctuations in the density of the universe. In our simulations, and in our
coordinates, the fluid is at rest throughout the simulation, so photons passing through
over-dense or under-dense regions will be redshifted according to how the metric has
responded to the over- or under-density. In synchronous gauge, this can be interpreted as
an analogue to a late-time integrated Sachs-Wolfe effect, or Rees-Sciama effect.
In a gauge where the fluid is not co-moving, at least some of this effect may interpreted
as due to peculiar velocities.
In Fig.~\ref{Hubble_overdense} we see such an effect for an
observer in a local over-density. We note that the effect of an observer sitting
in a local under-density is similar but opposite in magnitude and can mimic the
behavior of a cosmological constant. The idea that we are situated in a large
void has been explored in great detail in the past, see eg.
\cite{GarciaBellido:2008nz,Biswas:2010xm}. While such an effect is unlikely to
explain cosmic acceleration, it could nevertheless result in local measurements of
cosmological parameters that do not reflect more global properties of the
spacetime.

In a similar vein, we could use our technique to explore effects due to the
tendency of observable sources to lie in regions of higher density. One might
expect to see the average deviation from FLRW affected by a Sachs-Wolfe-type
effect, as photons not only climb out of gravitational potential wells,
but undergo additional Ricci focusing due to the presence of a local overdensity
at the source, resulting in a slightly diminished inferred angular diameter
distance. Preliminary explorations of this effect suggest this could
be an interesting future topic of study.

We conclude by examining the ability of our technique to explore differences
in inferred angular diameter distances for objects of large angular extent due
to inhomogenities. Such effects can manifest themselves in observations of
extended structures, for example degree-scale CMB fluctuations or
Baryon-Acoustic-Oscillation-scale structures. In principle, a similar
calculation can be used to directly probe breakdowns of the Etherington
Reciprocity theorem due to finite beam effects. Violations of this theorem are
of particular interest due to their ability to probe physics affecting the
underlying assumptions of the theorem, and have thus become increasingly well
constrained \cite{Rasanen:2015kca}.

As a simplistic experiment, consider two pointlike sources with a known
luminosity, part of a structure of known size, so the distance between
them is known. Thus the luminosity distance is known to each pointlike source
independently, and the angular diameter distance to the aggregate object can
be inferred. In the limit that the point sources are infinitesimally separated,
the relationship between the distance measures should obey Etherington's
reciprocity theorem. However as the source separation -- and therefore the
separation between rays -- increases, the inferred angular diameter distance
may no longer agree with the luminosity distances due to different lines of
sight probing different metric potentials. Thus, the inferred angular diameter
distance will disagree with at least one of the luminosity distances.

\begin{figure}[htb]
  \centering
    \includegraphics[width=0.45\textwidth]{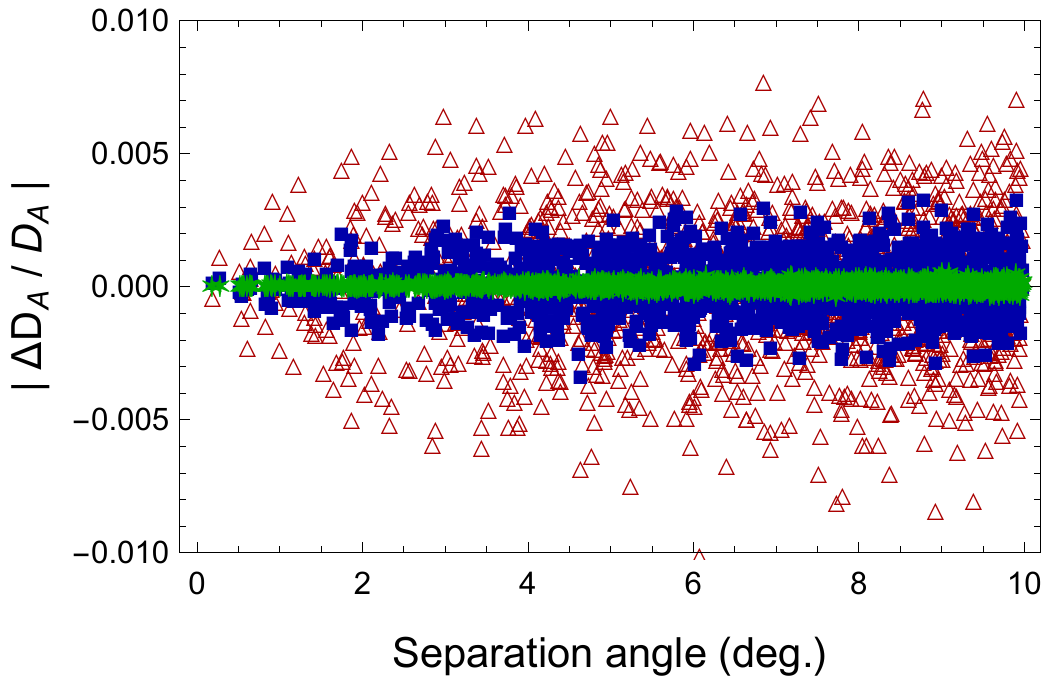}
  \caption{\label{DA_dist} Distribution of fractional differences in angular
  diameter distance, $\Delta D_A / D_A$, for infinitesimal beams separated by an
  angle. Beams are integrated out to sources at $z=0.1$ (green stars), $z=0.5$
  (blue squares), and $z=1.0$ (red triangles). At small separation, angular diameter
  distances agree; however at larger separation, increasing disagreement is found.}
\end{figure}

In Fig.~\ref{DA_dist}, we present the difference in angular diameter distances
computed by integrating the optical scalar equations, to an aggregate `source'
with finite angular extent, assuming the source is located at a fixed redshift.
We find very small deviations of distance measures on average and for beams with
small area, however individual measurements demonstrate appreciable deviations,
particularly on large angular scales. For actual observations, the situation may
differ in several regards. For example, if sources used as standard rulers lie
in similar gravitational potentials, the particular contribution of gravitational
redshift or peculiar velocity may be reduced.

\section{Conclusions and Discussion}

The results presented here are the next step in an important advancement of
numerical techniques that allows us to directly integrate the full Einstein
field equations and the optical scalar equations in an arbitrary spacetime. This
in turn enables computation of cosmological observables and directly measurable
quantities. As numerical relativity continues to find new applications in
cosmology, long-standing questions will be resolved in regimes where
non-linearities are significant. Determining the relevance of, and precisely
quantifying these effects in relativistic models will be an important future
task for this field.

Here we have examined the ability of techniques from numerical relativity to
probe aspects of cosmology where precise estimates are required. We consider
a universe with large-scale inhomogeneities, so while we expect perturbation
theory and approximation schemes such as Dyer-Roeder may accurately describe
the effects we observe, it is encouraging to see the well-understood physical
effects manifesting themselves in this model.

Numerous applications of numerical relativity can be made in future studies, and
it will be interesting to see the direction these studies take. For example,
while statistical studies can offer general insight into physical processes, it
may also be possible to model specific systems to better understand dynamics and
extract cosmological information.

\section{Acknowledgments}
We would like to thank Thomas Baumgarte for a series of very valuable
conversations, as well as Stuart Shaprio and Lorenzo Sorbo for conversations
that helped shape this work. We are also grateful for feedback from Pierre Fleury 
on an earlier draft of this manuscript.  JTG is supported by the National Science
Foundation, PHY-1414479; JBM and GDS are supported by a Department of Energy
grant DE-SC0009946 to CWRU. The simulations in this work made use of the
High Performance Computing Resource in the Core Facility for Advanced
Research Computing at Case Western Reserve University, and of hardware
provided by the National Science Foundation, the Research Corporation for
Science Advancement, and the Kenyon College Department of Physics.

\bibliography{references}

\begin{appendices}

\section{Angular Diameter Distances}
\label{angular_diameter_appendix}

The optical focusing equations \cite{Sachs:1961zz},
\begin{eqnarray}
\frac{{\rm d}}{{\rm d}\lambda}\theta+\theta^{2}+|\sigma|^{2} & = & \mathcal{R}\nonumber \\
\frac{{\rm d}}{{\rm d}\lambda}\sigma+2\theta\sigma & = & \mathcal{W}\,,\label{eq:scalarev}
\end{eqnarray}
describe the evolution of a beam with infinitesimal area. The standard variable
$\theta=1/2A \,{\rm d}A/{\rm d}\lambda$ describes the rate of expansion of a
beam's area $A$ (and is not an angle). The equations of motion we use, however,
work in terms of the rooted beam area $\ell=\sqrt{A}$, for which
$\theta=1/\ell \,\, {\rm d}\ell/{\rm d}\lambda$.
The quantity $\sigma$ is a complex scalar, loosely describing the shear rate
of a light beam up to additional geometric factors. The Weyl and Ricci optical
scalars, $\mathcal{W}$ and $\mathcal{R}$ respectively, are
\begin{eqnarray}
\mathcal{W} & = & -\frac{1}{2}C_{\mu\nu\rho\sigma}(s_{1}^{\mu}-is_{2}^{\mu})p^{\nu}p^{\rho}(s_{1}^{\sigma}-is_{2}^{\sigma})\\
\mathcal{R} & = & -4\pi T_{\mu\nu}p^{\mu}p^{\nu}=-\frac{1}{2}R_{\mu\nu}p^{\mu}p^{\nu} \label{eq:ricciopticalscalar}
\end{eqnarray}
for screen vectors $s_{1}^{\mu}$, $s_{2}^{\mu}$, and Weyl tensor $C_{\mu\nu\rho\sigma}$.
The Weyl scalar here is analogous to those from the Newman-Penrose
formalism for identifying gravitational radiation \cite{Newman:1961qr}; however
the vectors used to compute the Weyl scalar are comprised of a single null
vector $p^{\mu}$ and the screen vectors rather than a null tetrad.

The screen vectors obey the following orthogonality relationships:
$g_{\mu\nu}s_{N}^{\nu}p^{\mu}=0$ (normal to the photon 4-vector),
$g_{\mu\nu}s_{N}^{\nu}(q^\mu - U^\mu)\equiv g_{\mu\nu}s_{N}^{\nu}d^\mu=0$
(normal to the `direction of observation' $d^\mu$ and thus observer 4-velocity
$U^\mu$), and  $g_{\mu\nu}s_{1}^{\mu}s_{2}^{\nu}=0$, and
$g_{\mu\nu}s_{1}^{\mu}s_{1}^{\nu}=g_{\mu\nu}s_{2}^{\mu}s_{2}^{\nu}=1$
(orthonormality of $s_{1}^{\mu}$ and $s_{2}^{\mu}$). Additionally, they obey a
partial parallel transport equation \cite{Fleury:2015hgz},
\begin{equation}
P^{\mu}_{\nu}\frac{\rm{D}s_A^\nu}{\rm{d}\lambda} = 0
\end{equation}
for a screen projection operator
$P^{\mu\nu} = g^{\mu\nu} + U^\mu U^\nu - d^\mu d^\nu$. Written in 3+1 form in
synchronous gauge, the evolution equations for the screen vectors become
\begin{eqnarray}
\frac{\rm{d}}{\rm{d}t}s_A^\mu & = & s_{A}^{j}V^{i}\left(\gamma_{jk}\frac{d}{dt}V^{k}-3V^{k}K_{jk}+V_{l}V^{k}\Gamma_{kj}^{l}\right) \nonumber \\
 & & -\Gamma_{kl}^{i}V^{k}s_{A}^{l}+\gamma^{ij}K_{jk}s_{A}^{k} \,.
\end{eqnarray}
In practice, the term including the time derivative of $V^k$ is determined by
the left-hand side of Eq.~\ref{eq:Vev}. The computed beam area should be
invariant under rotations of the screen vectors in screen space, though not
invariant under time-dependent rotations. Therefore, a good test for code
validation is to vary the initial screen vectors and ensure there is no effect
on the optical scalars.

In practice, we substitute the definition of $\theta$ in terms of $\ell$, and
also define a variable analogous to the root solid angle subtended by the beam
according to an observer, $\varphi={\rm d}\ell/{\rm d}\lambda$
in Eq. \ref{eq:scalarev} (at an observer,
${\rm d}\ell_{obs}/{\rm d}\lambda=\sqrt{\Omega_{obs}}$,
where $\Omega_{obs}$ is the solid angle subtended by the beam). We write the
real and complex pieces of $\sigma$ separately, and also write the equations
in terms of coordinate time $t$. In order to further simplify the equations of
motion for $\sigma$, we define $\bar{\sigma}=\ell^{2}\sigma$.

Significantly, the evolution of the screen vectors is not only useful for
tracking the area of a beam; the polarization vectors of photons obey
an almost identical set of equations. Thus the equations we present here may
also provide a new technique for studying gravitational effects on polarization.
The major difference between evolution is only that normalization of screen
vectors is enforced throughout their evolution.

\section{Calculation of the Weyl scalar in Synchronous Gauge}
\label{weyl_calcs}

The dominant computational expense in integrating these equations through an
arbitrary spacetime will clearly come from computing Weyl tensor components.
In our code we employ synchronous gauge, where $\alpha=1$ and $\beta^{i}=0$.
There are in principle only ten independent components of the Weyl tensor
that need to be calculated; however in this work we will not take full
advantage of this, thus we compute 21 components of the Riemann tensor.
It may be interesting to study how alternative formulations compare, but we
do not pursue that idea in this work. In order to compute terms on the
right-hand side of the evolution equations, we first write
\begin{eqnarray}
\mathcal{W} & = & -\frac{1}{2}\left(R_{\mu\nu\rho\sigma}-(g_{\mu[\rho}R_{\sigma]\nu}-g_{\nu[\rho}R_{\sigma]\mu})\right.\nonumber\\
 & & \left.+\frac{1}{3}R\,g_{\mu[\rho}g_{\sigma]\nu}\right)(s_{1}^{\mu}-is_{2}^{\mu})p^{\nu}p^{\rho}(s_{1}^{\sigma}-is_{2}^{\sigma})\nonumber\\
 & = & -\frac{1}{2}R_{\mu\nu\rho\sigma}(s_{1}^{\mu}-is_{2}^{\mu})p^{\nu}p^{\rho}(s_{1}^{\sigma}-is_{2}^{\sigma})
\end{eqnarray}
as all contractions with $g_{\mu\nu}$ cancel and the last line follows
from $g_{\mu\nu}(s_{1}^{\mu}-is_{2}^{\mu})(s_{1}^{\nu}-is_{2}^{\nu})=1-1=0$.

In order to reduce this expression further, we will need to write the Riemann
tensor in terms of $3+1$ variables,

\begin{eqnarray}
^{(4)}R_{ilmj} & = & \frac{1}{2}\left(\gamma_{ij,lm}-\gamma_{jl,im}-\gamma_{im,lj}+\gamma_{ml,ij}\right)\nonumber\\
 & & +\Gamma_{qij}\Gamma_{lm}^{q}-\Gamma_{qim}\Gamma_{jl}^{q}-K_{im}K_{jl}+K_{ij}K_{ml} \nonumber\\
^{(4)}R_{i0mj} & = & \partial_{j}K_{im}-\partial_{m}K_{ij}+\left(\gamma_{qi,m}+\Gamma_{imq}\right)K_{j}^{q}\nonumber \\
 & & -\left(\gamma_{qi,j}+\Gamma_{ijq}\right)K_{m}^{q} \\
^{(4)}R_{i00j} & = & -^{(3)}R_{ij}-KK_{ij}+8\pi\left(S_{ij}-\frac{1}{2}\gamma_{ij}\left(S-\rho\right)\right) \nonumber
\end{eqnarray}

The term $\mathcal{W}$ can now be re-expressed so that a smaller number of
terms need actually be computed. Naively, there are $4\cdot4\cdot4\cdot4=256$
terms in the original expression,
\begin{equation}
R_{\alpha\mu\nu\beta}s_{\mathcal{A}}^{\alpha}k^{\mu}k^{\nu}s_{\mathcal{B}}^{\beta}\,,
\end{equation}
for $\mathcal{A},\,\mathcal{B}\in\{1,2\}$. To reduce the number of terms,
writing the index combination $\alpha,\mu\equiv A$ as a single index running
over all combinations of $\alpha$ and $\mu$, and denoting
$s_{\mathcal{A}}^{\alpha}k^{\mu}\equiv\varsigma_{\mathcal{A}}^{A}$
leads to some simplification:
\begin{eqnarray}
\mathcal{W} & \supset & -\frac{1}{2}R_{\alpha\mu\nu\beta}s_{\mathcal{A}}^{\alpha}k^{\mu}k^{\nu}s_{\mathcal{B}}^{\beta}	\nonumber \\
 & \equiv & \frac{1}{2}R_{AB}\varsigma_{\mathcal{A}}^{A}\varsigma_{\mathcal{B}}^{B}\\
 & \equiv & W_{\mathcal{AB}}^{\Sigma}\,.\nonumber
\end{eqnarray}
Because the indices $\alpha$ and $\mu$ are antisymmetric under interchange,
the indices $A$ and $B$ need not run over all values, but only 
\begin{equation}
A,\,B\in\{01,02,03,10,12,13,20,21,23,30,31,32\}\,.
\end{equation}
The antisymmetry further allows terms to be combined. Denoting the
canonical ordering of elements $A_{c}\in\{01,02,03,12,13,23\}$ and
reversed $A_{r}\in\{10,20,30,21,31,32\}$, so $A\in A_{c}\cup A_{r}$, 

\begin{eqnarray}
R_{AB}\varsigma_{\mathcal{A}}^{A}\varsigma_{\mathcal{B}}^{B} & = & R_{A_{c}B}\varsigma_{\mathcal{B}}^{B}\left(\varsigma_{\mathcal{A}}^{A_{c}}-\varsigma_{\mathcal{A}}^{A_{r}}\right)\\
 & \equiv & R_{A_{c}B}\varsigma_{\mathcal{B}}^{B}\left(2\varsigma_{\mathcal{A}}^{[A_{c}]}\right)
\end{eqnarray}
and similarly for $B$, so
\begin{equation}
R_{AB}\varsigma_{\mathcal{A}}^{A}\varsigma_{\mathcal{B}}^{B}=4R_{A_{c}B_{c}}\varsigma_{\mathcal{A}}^{[A_{c}]}\varsigma_{\mathcal{B}}^{[B_{c}]}\,.
\end{equation}
Because $A_{c}$ runs over 6 terms, and because $R_{AB}$ is symmetric, only $21$
total components of $R_{AB}$ need to be calculated in order to evaluate the
sum, along with the $6$ components in each of
$\varsigma_{\mathcal{A}}^{[A_{c}]}$ and $\varsigma_{\mathcal{B}}^{[B_{c}]}$.
The six $\varsigma_{\mathcal{A}}^{[A_{c}]}$ are

\begin{equation}
\begin{array}{ccccccc}
\varsigma_{\mathcal{A}}^{[01]} & = & -\frac{1}{2}s_{\mathcal{A}}^{1}p^{0} & \,\,\,\,\,\, & \varsigma_{\mathcal{A}}^{[12]} & = & \frac{1}{2}\left(s_{\mathcal{A}}^{1}p^{2}-s_{\mathcal{A}}^{2}p^{1}\right)\\
\varsigma_{\mathcal{A}}^{[02]} & = & -\frac{1}{2}s_{\mathcal{A}}^{2}p^{0} &  & \varsigma_{\mathcal{A}}^{[13]} & = & \frac{1}{2}\left(s_{\mathcal{A}}^{1}p^{3}-s_{\mathcal{A}}^{3}p^{1}\right)\\
\varsigma_{\mathcal{A}}^{[03]} & = & -\frac{1}{2}s_{\mathcal{A}}^{3}p^{0} &  & \varsigma_{\mathcal{A}}^{[23]} & = & \frac{1}{2}\left(s_{\mathcal{A}}^{2}p^{3}-s_{\mathcal{A}}^{3}p^{2}\right)
\end{array}
\end{equation}

and similarly for $\varsigma_{\mathcal{B}}^{[B_{c}]}$. The vectors $s^{\mu}$
can be chosen freely, so long as only the beam area and opening angle are of
interest, and they are chosen to be orthogonal to $p^{\mu}$ and the observer's
line of sight, $V^{i}$. The sum should then contain the $6\cdot6=36$ terms,

\begin{equation}
\begin{array}{cc}
 & \frac{1}{2}\mathcal{W}_{\mathcal{AB}}^{\Sigma}=\underset{A_{c},B_{c}\in\{01,02,03,12,13,23\}}{\sum}R_{A_{c}B_{c}}\varsigma_{\mathcal{A}}^{[A_{c}]}\varsigma_{\mathcal{B}}^{[B_{c}]}\,,\\
\end{array}
\end{equation}

which contains the 21 unique components of the Riemann tensor that can be
evaluated using Eq.\ref{eq:riemann_expansion}. The source terms for the optical
scalar equations are then written as
\begin{eqnarray}
\Re[\mathcal{W}] & = & \frac{1}{2}\left(\mathcal{W}_{11}^{\Sigma}-\mathcal{W}_{22}^{\Sigma}\right)\\
\Im[\mathcal{W}] & = & -\mathcal{W}_{12}^{\Sigma}\,.
\end{eqnarray}

\end{appendices}

\end{document}